\documentclass[aip,reprint]{revtex4-1}
\usepackage{graphicx}
\usepackage{dcolumn}
\usepackage{bm}
\usepackage{xcolor}

\begin{document}

\title{Sensitive capacitive pressure sensors based on graphene membrane arrays}

\author{Makars \v{S}i\v{s}kins}
\affiliation{Kavli Institute of Nanoscience, Delft University of Technology, Lorentzweg 1,
 2628 CJ, Delft, The Netherlands}
\author{Martin Lee}
\affiliation{Kavli Institute of Nanoscience, Delft University of Technology, Lorentzweg 1,
 2628 CJ, Delft, The Netherlands}
\author{Dominique Wehenkel}
\affiliation{Applied Nanolayers B.V., Feldmannweg 17, 2628 CT, Delft, The Netherlands}
\author{Richard van Rijn}
\affiliation{Applied Nanolayers B.V., Feldmannweg 17, 2628 CT, Delft, The Netherlands}
\author{Tijmen W. de Jong}
\affiliation{Kavli Institute of Nanoscience, Delft University of Technology, Lorentzweg 1,
 2628 CJ, Delft, The Netherlands}
\author{Johannes R. Renshof}
\affiliation{Kavli Institute of Nanoscience, Delft University of Technology, Lorentzweg 1,
 2628 CJ, Delft, The Netherlands}
\author{Berend C. Hopman}
\affiliation{Kavli Institute of Nanoscience, Delft University of Technology, Lorentzweg 1,
 2628 CJ, Delft, The Netherlands}
\author{Willemijn S. J. M. Peters}
\affiliation{Kavli Institute of Nanoscience, Delft University of Technology, Lorentzweg 1,
 2628 CJ, Delft, The Netherlands}
\author{Dejan Davidovikj}
\affiliation{Kavli Institute of Nanoscience, Delft University of Technology, Lorentzweg 1,
 2628 CJ, Delft, The Netherlands}
\author{Herre S. J. van der Zant}
\affiliation{Kavli Institute of Nanoscience, Delft University of Technology, Lorentzweg 1,
 2628 CJ, Delft, The Netherlands}
\author{Peter G. Steeneken}
\email[Electronic mail: ]{p.g.steeneken@tudelft.nl} 
\affiliation{Kavli Institute of Nanoscience, Delft University of Technology, Lorentzweg 1,
 2628 CJ, Delft, The Netherlands}
\affiliation{Department of Precision and Microsystems Engineering, Delft University of Technology,
	Mekelweg 2, 2628 CD, Delft, The Netherlands}

\date{\today}

\begin{abstract}
The high flexibility, impermeability and strength of graphene membranes are key properties that can enable the next generation of nanomechanical sensors. However, for capacitive pressure sensors the sensitivity offered by a single suspended graphene membrane is too small to compete with commercial sensors. Here, we realize highly sensitive capacitive pressure sensors consisting of arrays of nearly ten thousand small, freestanding double-layer graphene membranes. We fabricate large arrays of small diameter membranes using a procedure that maintains the superior material and mechanical properties of graphene, even after high-temperature anneals. These sensors are readout using a low cost battery-powered circuit board, with a responsivity of up to $47.8$ aF~Pa$^{-1}$mm$^{-2}$, thereby outperforming commercial sensors.
\end{abstract}

\maketitle
\section*{Introduction}
Graphene grown by chemical vapour deposition (CVD) is a strong candidate for realizing next-generation sensor devices \cite{Zurutuza2014}. Its hermeticity \cite{Lee2019, Bunch2008, SunGeim2019}, and superior mechanical \cite{Lee2008, Cui2020, Lee2013} and electrical \cite{Chen2008elect} properties, have enabled various types of gas pressure sensors. In most conventional pressure sensors, the pressure is determined from the deflection of a membrane due to the pressure difference between ambient gas and gas in a hermetically sealed reference cavity. There are different ways to readout the deflection of the membrane and thereby quantify the pressure difference. In capacitive graphene pressure sensors, the deflection is readout by measuring the capacitance between the graphene membrane and a reference electrode \cite{Chen2016capac,Berger2017,Berger2017touch, Davidovikj2017}. As the pressure induced deflection increases the mechanical stress and tension in the membrane, it can be measured using the piezoresistive effect \cite{Zhu2013,Smith2013, Smith2016piezo} and can also be probed via the mechanical resonance frequency \cite{Bunch2008, Patel2016,Lee2019}. In contrast, graphene squeeze-film pressure sensors \cite{Dolleman2015} and Pirani pressure sensors \cite{Romijn2018} do not require a hermetic reference cavity and operate at small deflection, which can be beneficial for their operation range.

Resonant sensors are conceptually attractive because they potentially offer both gas sensing \cite{Irek2020,Dolleman2016osmosis} and pressure sensing \cite{Lee2019, Dolleman2015} functionality within a single device. However, accurate readout of resonance frequencies with low-power electronics is challenging, requires elimination of mass loading and can not be easily scaled up to many devices in parallel. In piezoresistive and Pirani sensors, it is a challenge to eliminate non-pressure related effects of the surrounding gas on the graphene resistance. In contrast, capacitive pressure sensors have the advantage that the membrane capacitance is rather insensitive to gas induced changes in its mass and electrical resistance, and thus depends mainly on geometry and membrane deflection. However, it was found that a single graphene membrane with a diameter of 5 $\mu$m has a too small responsivity ($<0.1$ aF~Pa$^{-1}$) to be competitive with commercial sensors \cite{Davidovikj2017}.

In this work, we counter this drawback by creating arrays with a large number of membranes connected in parallel to increase the responsivity \cite{Davidovikj2017}. We present few-atom thick pressure sensors that can compete with commercial capacitive pressure sensors using arrays of nearly 10000 double-layer graphene (DLG) membranes. We optimize the design of the sensor elements, the chip layout and the readout electronics to attain a handheld, low-cost, battery-powered electrical readout circuit capable of detecting pressure changes via the static deflection of graphene nanodrums. 

\section*{Materials and Methods}
\subsection*{Chip design and graphene transfer}
\begin{figure}
	\includegraphics[scale=0.43]{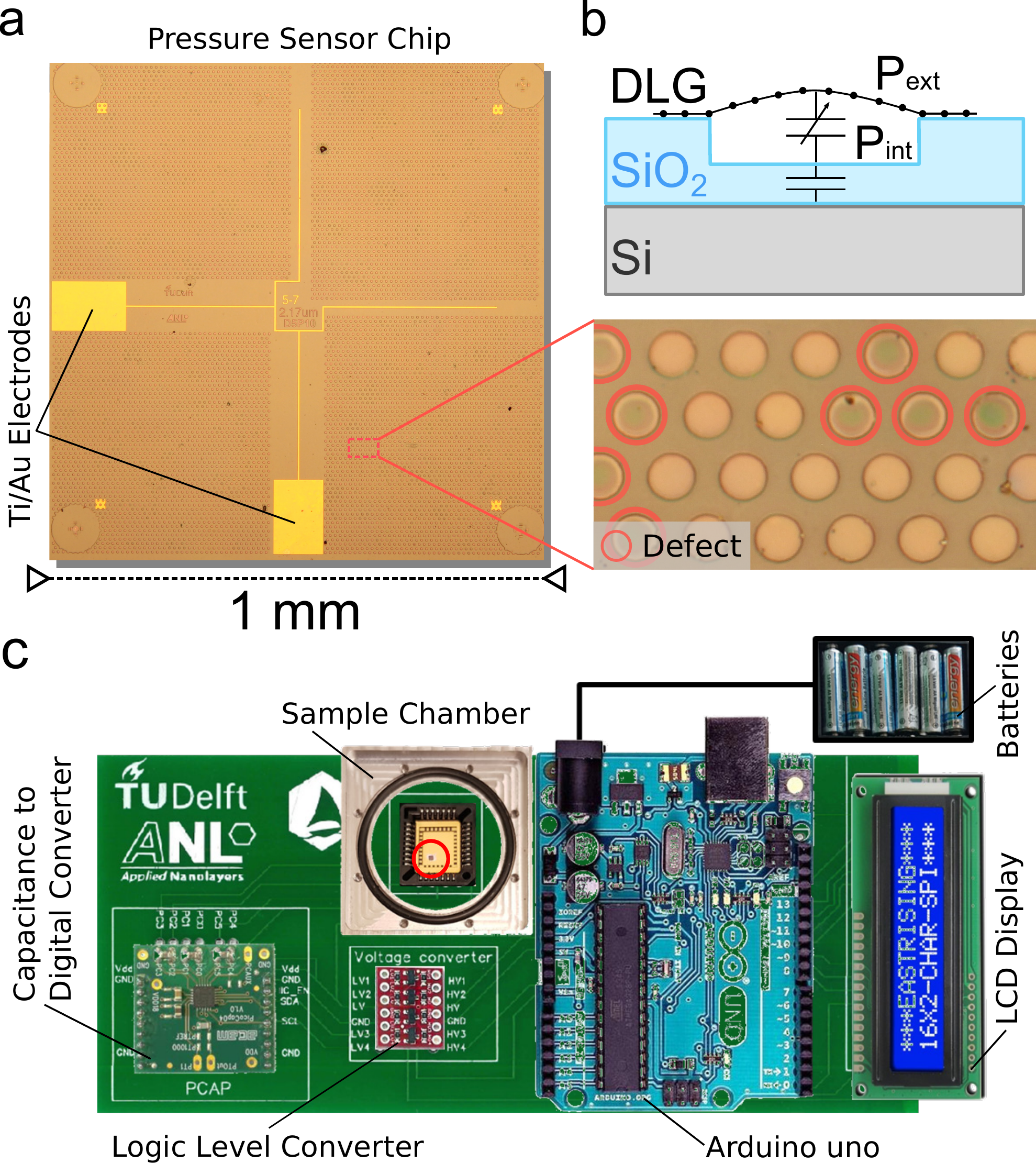}
	\caption{Pressure sensor and readout circuitry. (a) Optical image of the sensor chip with 10000 circular holes, gold electrodes and a DLG/PMMA membrane. Close-up image shows the difference in contrast between intact and defect drums with red circles indicating collapsed membranes. (b) Schematic device cross-section and capacitive pressure readout principle. (c) Read-out circuitry PCB board with the elements labeled. The red circle indicates the pressure sensor chip.}
	\label{fig:photo}
\end{figure}
Simulations \cite{Davidovikj2017} show that to achieve the commercially competitive sensitivity, an array of around 10000 circular graphene drums is needed, each with a diameter of $5$ micron. When the drums are placed on a hexagonal grid with a pitch of 10 micron between their centers, they fit on a $1\times 1$ mm$^2$ chip as shown in Fig. \ref{fig:photo}a. To fabricate this design, Ti/Au electrodes ($5$ nm/$60$ nm), for contacting the graphene top electrode, are patterned on a silicon chip with a $285$ nm SiO$_2$ layer. Then the pattern of circular holes with a depth of $240$ nm is reactive ion etched into the SiO$_2$. As shown in the cross-section in Fig.~\ref{fig:photo}b, the cavity depth of the holes is less than the SiO$_2$ layer thickness to prevent the graphene from touching the silicon bottom electrode, and thereby creating an electrical short-circuit between the electrodes, when one of the membranes collapses. As a last step the graphene is transferred over the cavities. We use two layers of graphene and small membrane diameter to improve the yield \cite{Cartamil2017} and the mechanical strength \cite{Barton2011, Lee2013}. Since the probability that $2$ pore defects align is low, the impermeability of DLG is also much higher than that of single layer graphene \cite{Bunch2008}.

To fabricate the double layer graphene (DLG), two sheets of CVD graphene are synthesized and then stacked on top of each other maintaining a pristine quality interface between the sheets. Using Polymethyl Methacrylate (PMMA) as a support layer of $800$ nm thickness, DLG is suspended over the pre-patterned circular holes in the SiO$_2$/Si chip with Ti/Au electrodes (Fig.~\ref{fig:photo}a). DLG was produced and transferred by Applied Nanolayers. From the differences in contrast between suspended, broken and collapsed drums \cite{CartamilBueno2016, Cartamil2017} we estimate that the dry transfer technique results in a yield of $95-99\%$ of freely suspended DLG/PMMA membranes. The red circles in the inset of Fig.~\ref{fig:photo}a indicate defect drums in a damaged region of the sample and show this difference in contrast.

\subsection*{Sensor readout circuit board}
\begin{figure}
	\includegraphics[scale=0.40]{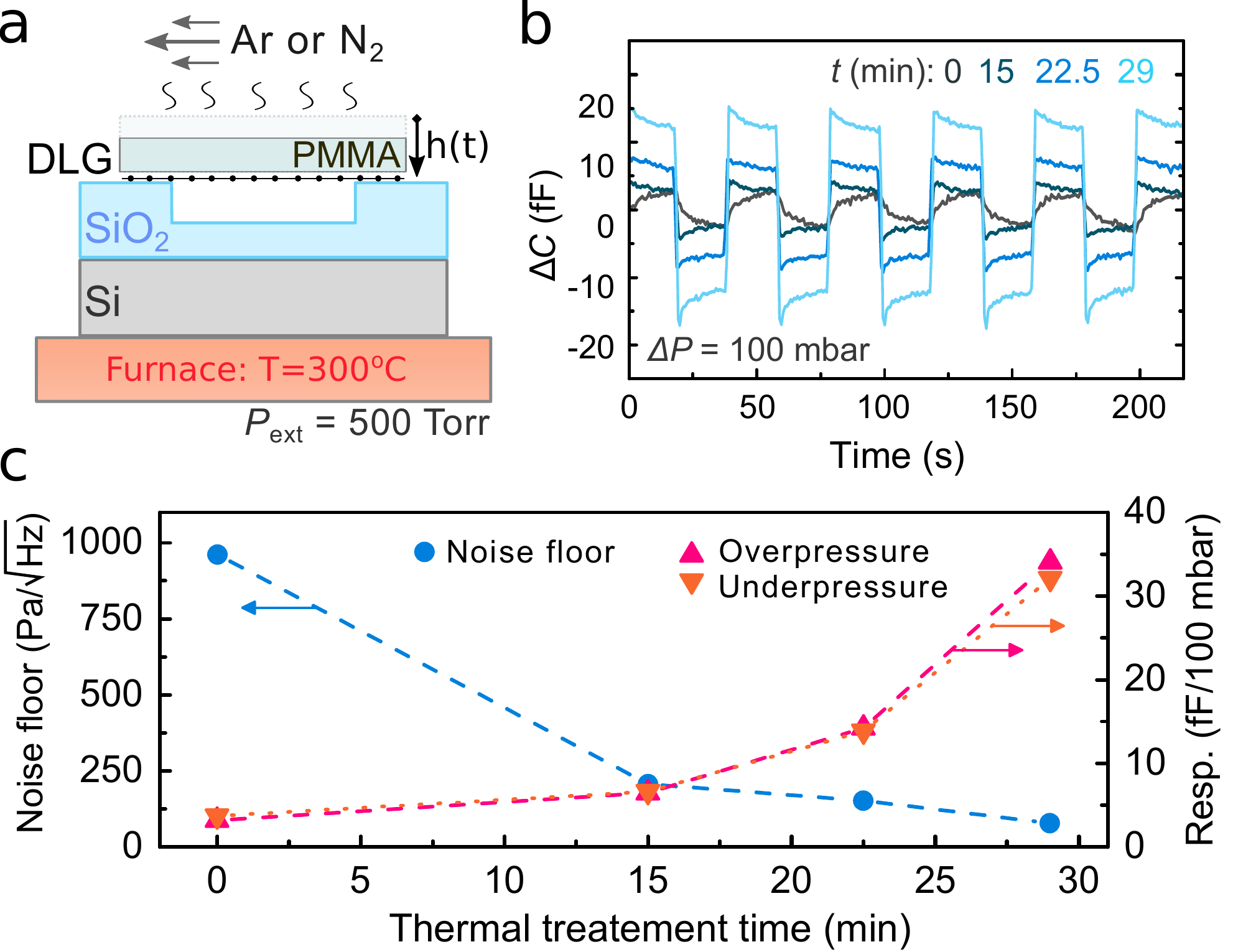}%
	\caption{Thermal removal of the polymer and its effect on the capacitive response of the sensor. (a) Schematic explanation of the thermal annealing principle that reduces the PMMA thickness $h(t)$. (b) Change in capacitance of a single chip as a function of measurement in response to time dependent pressure changes, $\Delta P$, for samples that have been annealed for different times $t$ (line colors corresponds to color of the corresponding anneal times indicated in the legend). (c) Responsivity and noise floor extracted from the data in (b) calculated using equation~(\ref{eq:noise}).}
	\label{baking}
\end{figure}
The graphene capacitive pressure sensor responds to the difference established between the internal pressure of the reference cavity, $P_{\rm{int}}$ and the external pressure of the environment, $P_{\rm{ext}}$. This pressure difference results in a deflection of the atomically thin membrane and a corresponding change in capacitance \cite{Davidovikj2017} of the drum, $\Delta C$ as schematically depicted in Fig.~\ref{fig:photo}b. The response of the graphene capacitive pressure sensor is high enough to be read-out by chip-scale commercial electronic components as is demonstrated using the battery-powered circuit shown in Fig.~\ref{fig:photo}c. Here, the sample under test is kept in a small on-board vacuum chamber that allows local control over the external gas pressure while preventing the pressure to affect electrical read-out elements. A commercial capacitance to digital converter (AMS PCap04) is then used to record and digitize the capacitance of the sample measured at a peak-to-peak voltage $V_{\rm{pp}}=1.5$ V with a hundred of charge and discharge cycles of $5.73$ ms each. After voltage level adjustment by a logic level converter, an Arduino Uno board converts the measured capacitance into pressure using a predetermined calibration curve and displays it on an LCD screen. The circuit board is capable of measuring a change in the chip's capacitance down to $\sim$10 aF on top of a background capacitance of a few tens of picofarads. 

\subsection*{Effect of PMMA removal}
After chip fabrication and transfer, the PMMA transfer polymer still covers the DLG sheets. In a number of previous studies protective polymer layers were used to support the graphene to increase the yield, mechanical performance, hermeticity and durability of the devices, in both suspended \cite{Berger2017, Berger2016sheets} and touch-mode operation \cite{Berger2017touch}. However, the bending rigidity of the polymer layers reduces the deflection and responsivity of the membranes and sensor. To improve the sensor performance we therefore gradually remove the transfer polymer by annealing it in dry gas \cite{burningLin2011,burningAhn2016,burningHuang2014,burningGammelgaard2014}, as shown in Fig.~\ref{baking}.

The sample is put inside a furnace and left at a pressure of $500$ Torr with a constant flow of an inert dry gas (Ar or N$_2$) at a temperature of $300^{\circ}$C as schematically depicted in Fig.~\ref{baking}a. We found no notable difference between the use of Ar or N$_2$ gas flow in terms of the end quality of DLG layers or PMMA removal rate. The average thickness of the transfer PMMA reduces with time at an estimated rate of $\sim27$ nm/minute with only minor residues left on and in-between the graphene layers. The thickness reduction by thermal treatment is found to substantially improve the responsivity of the sensors. Figure~\ref{baking}b shows the capacitance change $\Delta C$ of a single chip to external pressure changes between $1000$ and $900$ mbar with a period of 40 seconds as measured after $0$, $15$, $22.5$ and $29$ minutes of annealing. Figure~\ref{baking}c shows that by following the thermal anneal procedure, an increase in responsivity of almost an order of magnitude was achieved for this particular sample, while in the best case an increase in responsivity of nearly two orders of magnitude was observed after a $30$ minute continuous thermal anneal treatment. The detection noise floor also decreases substantially as a result of the process. The noise floor, $NF$, in Pa/$\sqrt{\rm{Hz}}$ is defined as:
\begin{equation}\label{eq:noise}
NF=\frac{C_{\rm{RMS}}}{\frac{d C}{d P}\sqrt{f_{\rm{meas}}}},
\end{equation}
where $C_{\rm{RMS}}$ is the RMS noise in the capacitance measurement, $\frac{ dC}{dP}$ the responsivity of the sensor and $f_{\rm{meas}}=1.745$ Hz, the frequency at which the measurements are acquired. The decrease in $NF$ as shown in Fig.~\ref{baking}c qualitatively follows from the equation~(\ref{eq:noise}) considering the measured increase in responsivity.

\begin{figure}
	\includegraphics[scale=0.49]{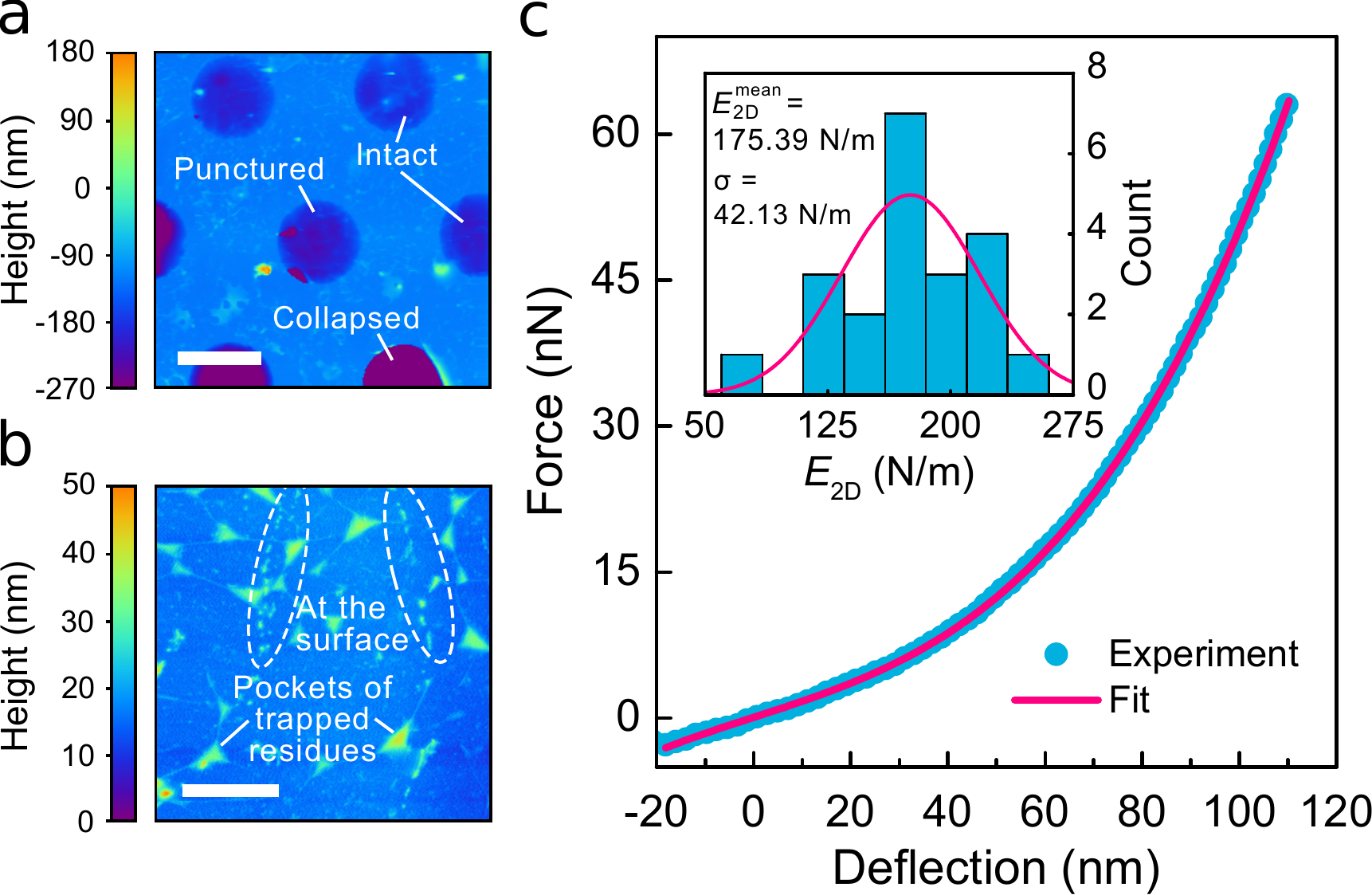}
	\caption{Atomic force microscopy (AFM) characterisation of membranes. (a) AFM tapping mode image of DLG drums. State of the drum is labeled. Scale bar: $5$ $\mu$m. (b) AFM tapping mode image of a supported region of DLG. Residue types are indicated. Scale bar: $1$ $\mu$m. (c) Force versus membrane deflection curve. Experimental data (blue dots) are fit by the membrane model of equation~(\ref{fdmodel}). Inset: statistics over 21 membranes with corresponding mean values for the extracted two-dimensional Young's modulus, $E_{\rm{2D}}$, with a mean pre-tension $n_0=0.04\pm0.02$ N/m.}
	\label{characterization}
\end{figure}

\section*{Results and Discussion}
\subsection*{Sample characterization}
After thermal treatment for 30 minutes at $300^{\circ}$C, we inspect the samples for damage. In Fig.~\ref{characterization}a a tapping mode Atomic Force Microscopy (AFM) image of the sample is shown. Three types of drums can be distinguished visually: intact, ruptured and collapsed drums. Collapsed drums are in contact with the bottom of the cavity and probably do not significantly contribute to the response \cite{Berger2017touch} because they are predominantly damaged and thus not airtight. The ruptured drums are also expected to leak fast \cite{Dolleman2015} and therefore have a negligible contribution to the static capacitance response to gradual pressure changes. Intact drums, however, show a full coverage of the cavity. These drums can hermetically seal the cavity with a constant internal pressure exploiting the extremely low permeability of graphene \cite{Bunch2008, Lee2019, SunGeim2019}, although in part of these membranes small pores can be present that are too small to be visually detected. As shown in the AFM measurements in Fig.~\ref{characterization}b, after the anneal, a substantial number of residues are observed on and below the DLG. The residues form pockets and are also observed in the suspended regions of the device, which suggests that part of the residues is trapped in-between the graphene layers, where they cannot be easily removed \cite{residuesJain2018}. 

\begin{figure}
	\includegraphics[scale=0.49]{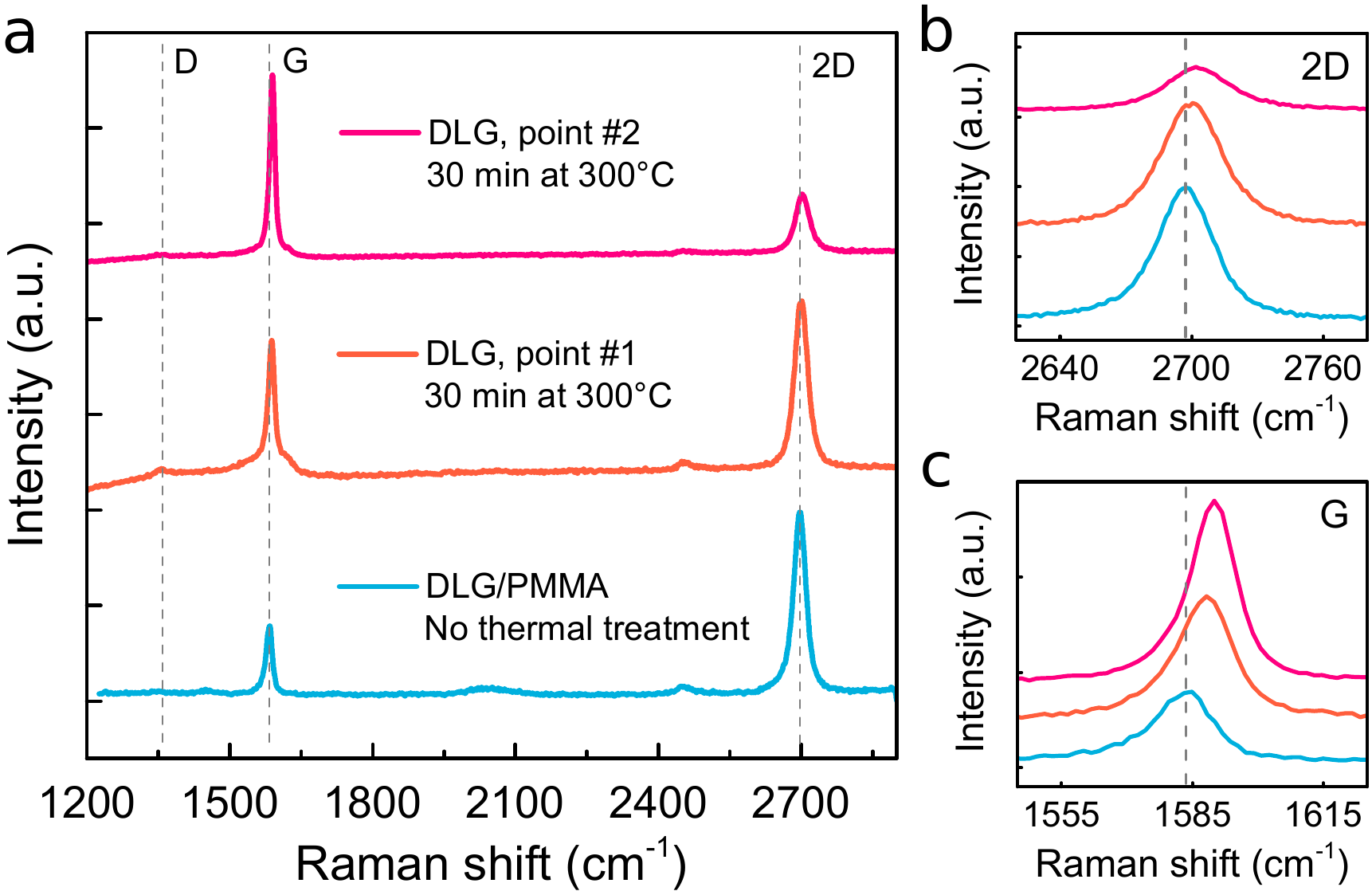}
	\caption{Raman spectroscopy of supported DLG. (a) Raman spectra of the DLG/PMMA layer before and after the polymer removal. (b) Close-up of a blue-shift in the 2D peak and (c) the G peaks of graphene.}
	\label{raman}
\end{figure}
During the fabrication, the capability of graphene to withstand high strains \cite{Lee2008,RuizVargas2011,Cui2020} facilitates damage-free transfer, while the thickness of only a few atoms favours higher membrane deflections and thus higher responsivity to pressure changes of the sensor. In order to test the elasticity and mechanical properties of the drums after thermal treatment, we use force-indention AFM to apply a point-force in the centre of a single membrane while measuring its deflection \cite{Lee2008}. The applied force, $F$, is proportional to the stiffness of a cantilever $k_c$ and its deflection $\Delta z_c$ as $F=k_c\Delta z_c$. We use two cantilevers of $k_c=1.25\pm0.12$ N/m and $1.48\pm0.05$ for two separate sets of measurements in two distant places on the chip. We record a force versus membrane deflection curve at the centre of each drum, as depicted in Fig.~\ref{characterization}c, and fit it to a model for point deflection of a circular membrane \cite{Lee2008, CastellanosGomez2012}: 
\begin{equation}\label{fdmodel}
F=n_0\pi \delta+E_{\rm{2D}}Rq^3\left(\frac{\delta}{R}\right)^3,
\end{equation}
where $n_0$ is the pre-tension, $E_{\rm{2D}}$ the two dimensional Young's modulus of the layer, $\nu=0.16$ the Poisson's ratio \cite{Lee2013}, $\delta$ the resulting deflection, $R$ the radius and $q=1/\left(1.05-0.15\nu-0.16\nu^2\right)$ a geometrical factor \cite{Lee2008, CastellanosGomez2012}. We use the two dimensional Young's modulus, $E_{\rm{2D}}=E_{\rm{3D}}h$ for the stacked DLG sheet since the thickness, $h$, of the layer after the thermal treatment is not well defined. Such an effective quasi-2D Young's modulus provides a more realistic estimate for the mechanical elasticity of the layer and can be directly compared to that of a pristine single layer graphene \cite{Lee2008}. In the inset of Fig.~\ref{characterization}c, the statistics over $21$ different drums are shown that yield a mean value of $E_{\rm{2D}}=175$ N/m. This is substantially lower than reported values for both exfoliated and pristine CVD single-layer graphene \cite{Lee2008,Lee2013}, but comparable to other CVD graphene membranes \cite{Nicholl2015,RuizVargas2011, Li2015, Berger2016sheets}, high-quality oxidized graphene sheets \cite{Suk2010,GmezNavarro2008} and other 2D materials like single-layer MoS$_2$ \cite{CastellanosGomez2012}. 
 
We also examined the sheets of DLG by using Raman spectroscopy, as displayed in Fig.~\ref{raman}. Figure~\ref{raman}a shows Raman spectra of DLG acquired before the removal of the PMMA layer (blue line) and after the processing at an elevated temperature (orange and magenta line). Before the thermal treatment, the Raman spectrum of DLG is reasonably homogeneous across the chip, showing a Lorenzian-shaped 2D peak of graphene and a well-defined G peak \cite{Ferrari2007}. Full width at half maximum (FWHM) of the 2D peak is around $\sim30$ cm$^{-1}$ and a high intensity ratio of 2D to G peaks resemble typical features of pristine graphene \cite{Ferrari2007}. In the case of DLG, this indicates that the two layers in the stack are well decoupled and/or have on average a twist-angle \cite{Kim2012} larger than $15^{\circ}$. After processing at a high temperature, the width of the 2D peak remains the same (see Fig.~\ref{raman}b) while the ratio between 2D to G peaks changes drastically depending on the chosen location of the measurement on the chip (see Fig.~\ref{raman}a, orange and magenta lines). Also, a notable blue-shift of both 2D and G peaks is observed, as shown in Fig.~\ref{raman}b,c. These observations are attributed to a substantial difference in twist-angle across the DLG sheet \cite{Kim2012} as well as local changes in strain as a result of annealing of the graphene layers \cite{Ni2008}. 

We also note the almost complete absence of the D peak in all Raman spectra as shown in Fig.~\ref{raman}a, indicating a very low amount of defects in the stacked graphene layers even after exposure to high temperatures \cite{Eckmann2012}. This result is in agreement with the outstanding high-temperature stability of graphene when encapsulated by protective layers \cite{siskins2019, Son2017}, and provides evidence that damage, caused by the removal of polymer from suspended graphene, is minimal \cite{burningLin2011,burningAhn2016,burningHuang2014,burningGammelgaard2014}.

\begin{figure}
	\includegraphics[scale=0.20]{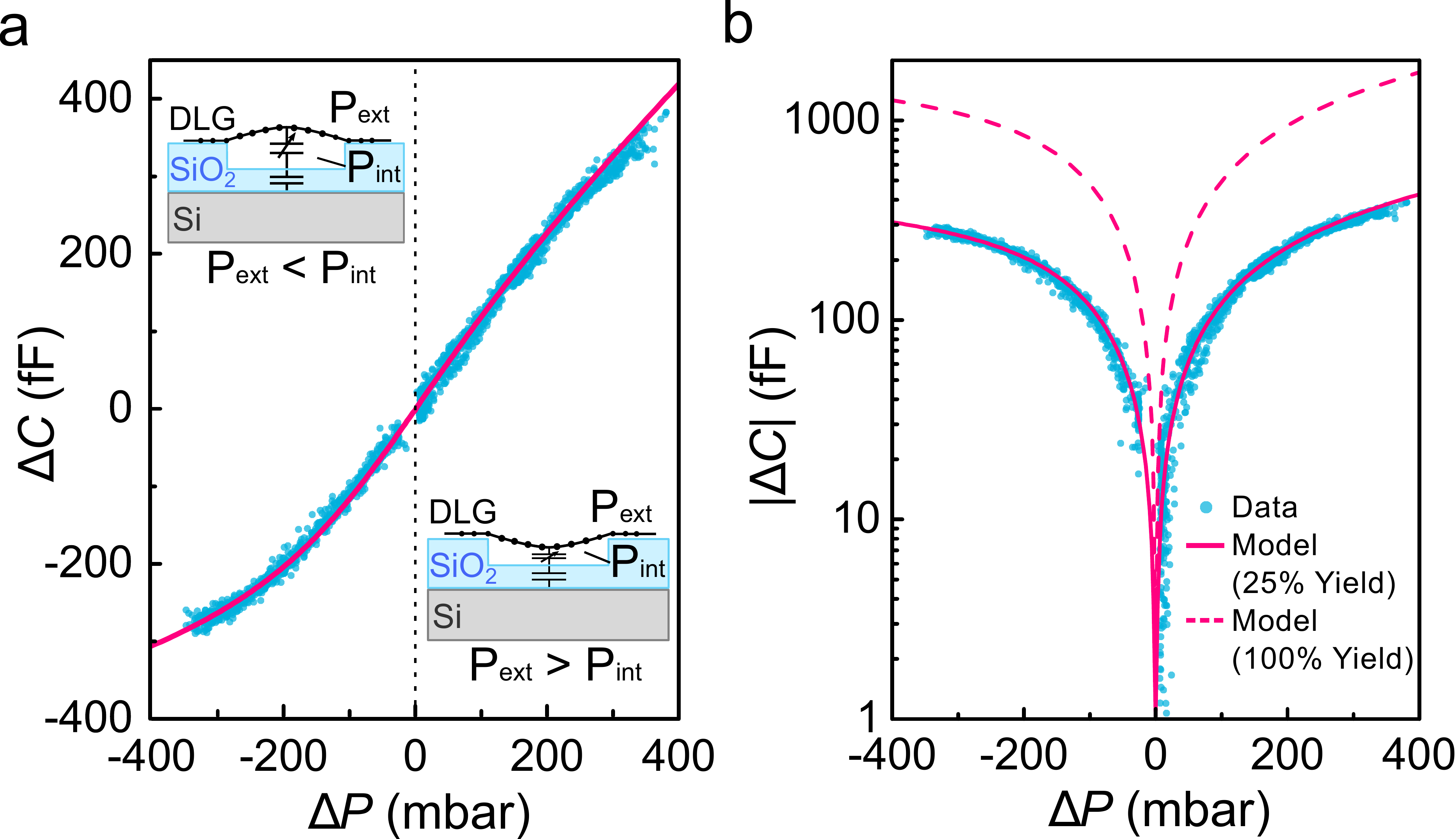}
	\caption{Capacitance-pressure curves of the DLG pressure sensor. (a) Measured capacitance-pressure curve (blue points); solid magenta line is a fit of equation (\ref{eq:dev}) to the data with $N/N_{tot}$=0.25, $E_{2D}$=175 N/m and $n_0=0.45$ N/m. Insets: schematic images of the effect of membrane deflection. (b) Comparison, on log-scale, of the measured data (blue points) to the model with $100\%$ (magenta dashed line) and $25\%$ (solid magenta line) yield of the hermetic drums using the 2D Young's modulus from Fig.~\ref{characterization}. All data is acquired during a continuous measurement within a total time of $1$ hour at $P_{\rm{ext}} \neq P_{\rm{int}}$.}
	\label{aligned}
\end{figure}
\begin{figure*}
	\includegraphics[scale=0.435]{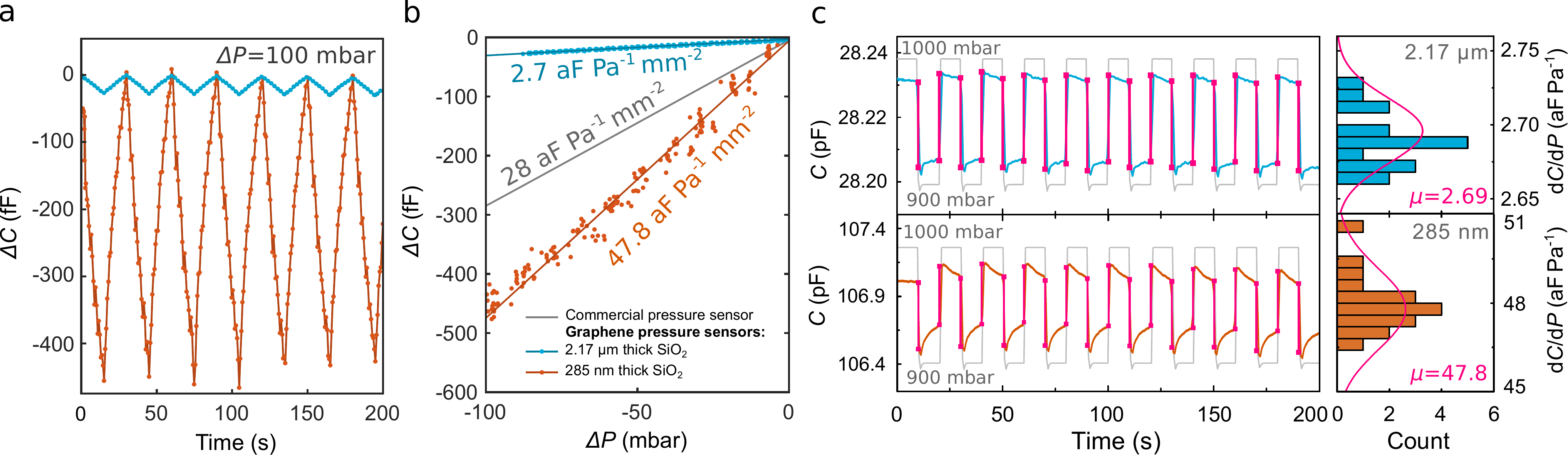}
    	\caption{Comparison of different capacitive pressure sensors.  (a) Triangular pressure wave applied to two graphene sensors with $285$ nm and $2.17$ $\mu$m of oxide thickness showing the measured change in capacitance. (b) Responsivity of the chips from (a) at small $\Delta P$ and the comparison to a commercial capacitive pressure sensor (Murata SCB10H). (c) Analysis of the responsivity of the chips as determined from periodic pressure steps between $1000$ and $900$ mbar. Panels on the left: Blue line - measured capacitance of a graphene sensor with $t_{SiO_2}=2.17$ $\mu$m. Orange line - measured capacitance of a sensor with $t_{SiO_2}=285$ nm. Thin grey line - applied time-dependent pressure profile. Magenta lines - extracted capacitance response of the sensors. Panels on the right: statistics over a number of cycles for both oxide thicknesses. Magenta lines - fit to a normal distribution with corresponding mean values $\mu$ in aF~Pa$^{-1}$ indicated.} 
	\label{cap}
\end{figure*}
\subsection*{Analysis of the sensor response}
A pressure difference $\Delta P$ results in a deflection $\delta$ of a circular graphene membrane with radius $R$, given by: 
\begin{equation}\label{eq:spherical}
    \Delta P=\frac{4n_0}{R^2}\delta+\frac{8E_{\rm{2D}}}{3R^4(1-\nu)}\delta^3,
\end{equation}
where the graphene membrane takes the shape of the section of a sphere \cite{bunch2008mechanical, Davidovikj2017}. Since the pressure inside the reference cavity is about $P_{int}\approx 1$ bar (the pressure during transfer), at $P_{ext}=1$ bar atmospheric pressure $\Delta P\approx 0$ and according to equation~(\ref{eq:spherical}) the sensors are expected to have a linear response at small $\Delta \delta$. However, at larger deflections nonlinear mechanical and capacitance effects start to result in non-linearities in the capacitance $C_{\rm{d}}(\Delta P)$ curve that can be calculated using the parallel-plate approximation \cite{Davidovikj2017} as:
\begin{equation}\label{eq:cap}
C_{\rm{d}}(\Delta P)=2 \pi \epsilon_0 \int_{0}^{R} \frac{r}{g_0-\delta(\Delta P) \left(1-\frac{r^2}{R^2}\right)} \rm{d} r,
\end{equation}
where $\epsilon_0$ is the vacuum permittivity, $g_0$ the gap size between the membrane and bottom electrode for $\Delta P = 0$. The contribution of quantum capacitance of graphene is small \cite{qcPonomarenko2010,qcXia2009} and is neglected. The total capacitance change of the sensor can be modeled from equation~(\ref{eq:spherical}) and (\ref{eq:cap}) as:
\begin{equation}\label{eq:dev}
    \Delta C_{\rm{total}}=N\times\Delta C_{\rm{d}}(\Delta P), 
\end{equation}
where $N$ is the number of intact, hermetic drums after both fabrication and thermal treatment.  
We experimentally test if equations~(\ref{eq:cap}) and (\ref{eq:dev}) can model the graphene pressure sensor by applying both substantial negative and positive pressure differences while measuring its capacitance. Figure~\ref{aligned}a shows the response of the same sensor that was characterized in Figs.~\ref{characterization} and~\ref{raman}.

The maximal responsivity of the sensor is achieved near ambient pressure in the linear regime, while a notable nonlinear response occurs for $|\Delta P|>200$ mbar. A number of design factors, such as the drum diameter, the number of drums and the pitch between the drums influence the sensor performance \cite{Davidovikj2017}. Importantly, the pressure dependence of $\Delta C$ can be well reproduced by the model of equation~(\ref{eq:dev}) using a Young's modulus of $E_{\rm{2D}} = 175$ N/m, estimated by an AFM probe for this particular sample, and a pre-tension $n_0=0.45$ N/m as shown by the magenta line in Fig.~\ref{aligned}a. The model follows the measurement closely when we use a fitted value of $N/N_{tot}$=0.25 as shown in Fig.~\ref{aligned}b (solid magenta line), where $N_{tot}\approx 10000$ is the total number of drums. This indicates that the yield of intact hermetic drums is $25\%$, and suggests that a large number of drums that look visually intact do not remain hermetic after polymer removal. The theoretical maximum response at a perfect yield of $100\%$ is also shown (dashed magenta line).

Many other factors can influence the responsivity. Since the device capacitance has a strong dependence on the distance between the plates of the pressure sensor, the cavity depth has a large influence on the performance of the sensor. In Fig.~\ref{cap} we demonstrate the performance of two of the best samples with a SiO$_2$ thickness of $285$ nm and $2.17$ $\mu$m respectively. Both chips have circular holes with a depth of $240$ nm fabricated as described previously. Figure~\ref{cap}a shows the triangular wave response of both $285$ nm (orange line) and $2.17$ $\mu$m (blue line) chips. As shown in Fig.~\ref{cap}b, the difference in responsivity for the two oxide thicknesses as a function of pressure can be more than an order of magnitude. Using equation~(\ref{eq:noise}), we calculate the noise level to be $34.2$ Pa/$\sqrt{\rm{Hz}}$ for the $285$ nm thick SiO$_2$ sample and $43.4$ Pa/$\sqrt{\rm{Hz}}$ for the $2.17$ $\mu$m one. 

There is a notable scatter in the measured capacitance values as seen in Fig.~\ref{cap}b that is attributed to the effect of gas escaping the cavities for part of the drums, thus causing $|\Delta P|$ to decrease with time. It has been reported before that when graphene is suspended over a SiO$_2$/Si cavity, it does not always form a perfect hermetic seal\cite{Bunch2008, Lee2019}. As recently shown by Lee, \textit{et al.}, most of the gas permeation in graphene drums occurs along the van der Waals interface between the 2D material and the substrate \cite{Lee2019}. As a result of the contribution of this effect, an exponential decrease of $C$ with time is visible in Fig.~\ref{cap}c, in response to periodic pressure steps of $\Delta P=100$ mbar. However, if a good graphene crystallinity is preserved over large areas, the mean path that the gas needs to travel to escape the cavities becomes large, and this will increase the flow resistance of the gas channel and the permeation time constant \cite{Lee2019,Dolleman2016osmosis}. This condition seems to be achieved across part of the sample, because no significant hysteresis was observed during the slow pressure sweeps ($>10$ minutes per sweep) in Fig.~\ref{aligned}a and b, suggesting that of the order of 25 \% of the drums maintain a constant internal pressure $P_{\rm{int}}$ for $\Delta P \neq 0$, as supported by the fits in Fig.~\ref{aligned}a and b.

Optimization of all parameters is required to achieve the best sensor performance for detecting the smallest pressure differences. In terms of responsivity our best sensor with $285$ nm of SiO$_2$ oxide (orange dots in Fig.~\ref{cap}a and b), with a responsivity of $47.8$ aF~Pa$^{-1}$mm$^{-2}$, already outperforms the commercially available state-of-the-art Murata SCB10H sensor with a responsivity of $28$ aF~Pa$^{-1}$mm$^{-2}$, as shown in Fig.~\ref{cap}b. This is comparably larger than what was previously achieved with capacitive sensors based on atomically thin 2D membranes \cite{Davidovikj2017, Chen2016capac}, yet competitive with a thicker suspended graphene-covered $140$ nm PMMA sheets \cite{Berger2017} with a reported responsivity of $123$ aF~Pa$^{-1}$mm$^{-2}$ over $210$ $\mu$m$^2$ of tightly-packed hexagonal membranes. However, if yields close to $100\%$ and hermetic sealing of all drums is realized and the pre-tension is lowered down to $\sim0.04$ N/m, the presented graphene-based 2D devices are expected to achieve a theoretical maximum responsivity of $\sim450$ aF~Pa$^{-1}$mm$^{-2}$ near ambient pressure. Further improvement might even be possible if the gap is reduced and the packing density of the membranes is increased.

\section*{Conclusions}
Where previous studies had addressed part of the challenges related to realizing portable graphene gas pressure sensors, like the pressure sensitivity, impermeability to gasses and electrical readout, here we bridged these studies by constructing a portable, battery powered functional graphene pressure sensor that outperforms commercial devices. Using off-the-shelf systems for electronic readout and data processing, we enable capacitive readout of a $1\times1$ mm$^2$ array of DLG graphene pressure sensors. We realize sensor chips with a high yield of suspended membranes, resulting in a sensor responsivity of $47.8$ aF~Pa$^{-1}$mm$^{-2}$. We demonstrate that thermal treatment is an effective measure for controllable thickness reduction of the support polymer layer, that leads to a significant performance improvement because the thin 2D material membranes are much more flexible than with the support polymer. It is anticipated that further design and fabrication improvements and better control over the device yield can increase the responsivity by a factor 10, thus enabling improvements in applications like indoor navigation, altitude monitoring and activity monitoring, and can enable new applications like presence detection.

\begin{acknowledgments}
M.\v{S}., M.L., H.S.J.v.d.Z. and P.G.S. acknowledge funding from the European Union’s Horizon 2020 research and innovation program under grant agreement number 785219. 
\end{acknowledgments}

\section*{Author contributions}
M.L. and D.D. designed and fabricated Si/SiO$_2$ chips. R.v.R. and D.W. fabricated and transferred the CVD graphene stack at Applied Nanolayers B.V. M.\v{S}. and M.L. realised a thermal removal of the polymer. T.W.d.J., J.R.R. and D.D. designed the read-out circuit board. M.\v{S}., M.L., T.W.d.J., J.R.R., B.C.H., W.S.J.M.P. and D.D. performed capacitance measurements. T.W.d.J., J.R.R., B.C.H. and D.D. analyzed the noise level of the sensors. W.S.J.M.P. and M.\v{S}. performed the AFM measurements. W.S.J.M.P., M.\v{S}. and M.L. modeled a response of the sensor. D.D., H.S.J.v.d.Z. and P.G.S. conceived and supervised the project. The manuscript was jointly written by all authors with a main contribution from M.\v{S}. All authors discussed the results and commented on the manuscript.

\section*{Conflict of interest}
The authors declare that they have no conflict of interest.


\section*{References}
\begin{thebibliography}{10}
\expandafter\ifx\csname url\endcsname\relax
  \def\url#1{\texttt{#1}}\fi
\expandafter\ifx\csname urlprefix\endcsname\relax\def\urlprefix{URL }\fi
\providecommand{\bibinfo}[2]{#2}
\providecommand{\eprint}[2][]{\url{#2}}

\bibitem{Zurutuza2014}
\bibinfo{author}{Zurutuza, A.} \& \bibinfo{author}{Marinelli, C.}
\newblock \bibinfo{title}{Challenges and opportunities in graphene
  commercialization}.
\newblock \emph{\bibinfo{journal}{Nature Nanotech.}}
  \textbf{\bibinfo{volume}{9}}, \bibinfo{pages}{730--734}
  (\bibinfo{year}{2014}).

\bibitem{Lee2019}
\bibinfo{author}{Lee, M.} \emph{et~al.}
\newblock \bibinfo{title}{Sealing graphene nanodrums}.
\newblock \emph{\bibinfo{journal}{Nano Lett.}} \textbf{\bibinfo{volume}{19}},
  \bibinfo{pages}{5313--5318} (\bibinfo{year}{2019}).

\bibitem{Bunch2008}
\bibinfo{author}{Bunch, J.~S.} \emph{et~al.}
\newblock \bibinfo{title}{Impermeable atomic membranes from graphene sheets}.
\newblock \emph{\bibinfo{journal}{Nano Lett.}} \textbf{\bibinfo{volume}{8}},
  \bibinfo{pages}{2458--2462} (\bibinfo{year}{2008}).

\bibitem{SunGeim2019}
\bibinfo{author}{Sun, P.~Z.} \emph{et~al.}
\newblock \bibinfo{title}{Limits on gas impermeability of graphene}.
\newblock \emph{\bibinfo{journal}{Nature}} \textbf{\bibinfo{volume}{579}},
  \bibinfo{pages}{229--232} (\bibinfo{year}{2020}).

\bibitem{Lee2008}
\bibinfo{author}{Lee, C.}, \bibinfo{author}{Wei, X.}, \bibinfo{author}{Kysar,
  J.~W.} \& \bibinfo{author}{Hone, J.}
\newblock \bibinfo{title}{Measurement of the elastic properties and intrinsic
  strength of monolayer graphene}.
\newblock \emph{\bibinfo{journal}{Science}} \textbf{\bibinfo{volume}{321}},
  \bibinfo{pages}{385--388} (\bibinfo{year}{2008}).

\bibitem{Cui2020}
\bibinfo{author}{Cui, T.} \emph{et~al.}
\newblock \bibinfo{title}{Fatigue of graphene}.
\newblock \emph{\bibinfo{journal}{Nature Mater.}}  (\bibinfo{year}{2020}).

\bibitem{Lee2013}
\bibinfo{author}{Lee, G.-H.} \emph{et~al.}
\newblock \bibinfo{title}{High-strength chemical-vapor-deposited graphene and
  grain boundaries}.
\newblock \emph{\bibinfo{journal}{Science}} \textbf{\bibinfo{volume}{340}},
  \bibinfo{pages}{1073--1076} (\bibinfo{year}{2013}).

\bibitem{Chen2008elect}
\bibinfo{author}{Chen, J.-H.}, \bibinfo{author}{Jang, C.},
  \bibinfo{author}{Xiao, S.}, \bibinfo{author}{Ishigami, M.} \&
  \bibinfo{author}{Fuhrer, M.~S.}
\newblock \bibinfo{title}{Intrinsic and extrinsic performance limits of
  graphene devices on {SiO}2}.
\newblock \emph{\bibinfo{journal}{Nature Nanotech.}}
  \textbf{\bibinfo{volume}{3}}, \bibinfo{pages}{206--209}
  (\bibinfo{year}{2008}).

\bibitem{Chen2016capac}
\bibinfo{author}{Chen, Y.-M.} \emph{et~al.}
\newblock \bibinfo{title}{Ultra-large suspended graphene as a highly elastic
  membrane for capacitive pressure sensors}.
\newblock \emph{\bibinfo{journal}{Nanoscale}} \textbf{\bibinfo{volume}{8}},
  \bibinfo{pages}{3555--3564} (\bibinfo{year}{2016}).

\bibitem{Berger2017}
\bibinfo{author}{Berger, C.}, \bibinfo{author}{Phillips, R.},
  \bibinfo{author}{Centeno, A.}, \bibinfo{author}{Zurutuza, A.} \&
  \bibinfo{author}{Vijayaraghavan, A.}
\newblock \bibinfo{title}{Capacitive pressure sensing with suspended
  graphene{\textendash}polymer heterostructure membranes}.
\newblock \emph{\bibinfo{journal}{Nanoscale}} \textbf{\bibinfo{volume}{9}},
  \bibinfo{pages}{17439--17449} (\bibinfo{year}{2017}).

\bibitem{Berger2017touch}
\bibinfo{author}{Berger, C.} \emph{et~al.}
\newblock \bibinfo{title}{Touch-mode capacitive pressure sensor with
  graphene-polymer heterostructure membrane}.
\newblock \emph{\bibinfo{journal}{2D Mater.}} \textbf{\bibinfo{volume}{5}},
  \bibinfo{pages}{015025} (\bibinfo{year}{2017}).

\bibitem{Davidovikj2017}
\bibinfo{author}{Davidovikj, D.}, \bibinfo{author}{Scheepers, P.~H.},
  \bibinfo{author}{van~der Zant, H. S.~J.} \& \bibinfo{author}{Steeneken,
  P.~G.}
\newblock \bibinfo{title}{Static capacitive pressure sensing using a single
  graphene drum}.
\newblock \emph{\bibinfo{journal}{ACS Appl. Mater. Interfaces}}
  \textbf{\bibinfo{volume}{9}}, \bibinfo{pages}{43205--43210}
  (\bibinfo{year}{2017}).

\bibitem{Zhu2013}
\bibinfo{author}{Zhu, S.-E.}, \bibinfo{author}{Ghatkesar, M.~K.},
  \bibinfo{author}{Zhang, C.} \& \bibinfo{author}{Janssen, G. C. A.~M.}
\newblock \bibinfo{title}{Graphene based piezoresistive pressure sensor}.
\newblock \emph{\bibinfo{journal}{Appl. Phys. Lett.}}
  \textbf{\bibinfo{volume}{102}}, \bibinfo{pages}{161904}
  (\bibinfo{year}{2013}).

\bibitem{Smith2013}
\bibinfo{author}{Smith, A.~D.} \emph{et~al.}
\newblock \bibinfo{title}{Electromechanical piezoresistive sensing in suspended
  graphene membranes}.
\newblock \emph{\bibinfo{journal}{Nano Lett.}} \textbf{\bibinfo{volume}{13}},
  \bibinfo{pages}{3237--3242} (\bibinfo{year}{2013}).

\bibitem{Smith2016piezo}
\bibinfo{author}{Smith, A.~D.} \emph{et~al.}
\newblock \bibinfo{title}{Piezoresistive properties of suspended graphene
  membranes under uniaxial and biaxial strain in nanoelectromechanical pressure
  sensors}.
\newblock \emph{\bibinfo{journal}{{ACS} Nano}} \textbf{\bibinfo{volume}{10}},
  \bibinfo{pages}{9879--9886} (\bibinfo{year}{2016}).

\bibitem{Patel2016}
\bibinfo{author}{Patel, R.~N.}, \bibinfo{author}{Mathew, J.~P.},
  \bibinfo{author}{Borah, A.} \& \bibinfo{author}{Deshmukh, M.~M.}
\newblock \bibinfo{title}{Low tension graphene drums for electromechanical
  pressure sensing}.
\newblock \emph{\bibinfo{journal}{2D Mater.}} \textbf{\bibinfo{volume}{3}},
  \bibinfo{pages}{011003} (\bibinfo{year}{2016}).

\bibitem{Dolleman2015}
\bibinfo{author}{Dolleman, R.~J.}, \bibinfo{author}{Davidovikj, D.},
  \bibinfo{author}{Cartamil-Bueno, S.~J.}, \bibinfo{author}{van~der Zant, H.
  S.~J.} \& \bibinfo{author}{Steeneken, P.~G.}
\newblock \bibinfo{title}{Graphene squeeze-film pressure sensors}.
\newblock \emph{\bibinfo{journal}{Nano Lett.}} \textbf{\bibinfo{volume}{16}},
  \bibinfo{pages}{568--571} (\bibinfo{year}{2015}).

\bibitem{Romijn2018}
\bibinfo{author}{Romijn, J.} \emph{et~al.}
\newblock \bibinfo{title}{A miniaturized low power pirani pressure sensor based
  on suspended graphene}.
\newblock In \emph{\bibinfo{booktitle}{2018 {IEEE} 13th Annual International
  Conference on Nano/Micro Engineered and Molecular Systems ({NEMS})}}
  (\bibinfo{publisher}{{IEEE}}, \bibinfo{year}{2018}).

\bibitem{Irek2020}
\bibinfo{author}{Rosło\'{n}, I.~E.} \emph{et~al.}
\newblock \bibinfo{title}{Graphene effusion-based gas sensor}
  (\bibinfo{year}{2020}).
\newblock \bibinfo{note}{Preprint at https://arxiv.org/abs/2001.09509v1}.

\bibitem{Dolleman2016osmosis}
\bibinfo{author}{Dolleman, R.~J.}, \bibinfo{author}{Cartamil-Bueno, S.~J.},
  \bibinfo{author}{van~der Zant, H. S.~J.} \& \bibinfo{author}{Steeneken,
  P.~G.}
\newblock \bibinfo{title}{Graphene gas osmometers}.
\newblock \emph{\bibinfo{journal}{2D Mater.}} \textbf{\bibinfo{volume}{4}},
  \bibinfo{pages}{011002} (\bibinfo{year}{2016}).

\bibitem{Cartamil2017}
\bibinfo{author}{Cartamil-Bueno, S.~J.} \emph{et~al.}
\newblock \bibinfo{title}{Very large scale characterization of graphene
  mechanical devices using a colorimetry technique}.
\newblock \emph{\bibinfo{journal}{Nanoscale}} \textbf{\bibinfo{volume}{9}},
  \bibinfo{pages}{7559--7564} (\bibinfo{year}{2017}).

\bibitem{Barton2011}
\bibinfo{author}{Barton, R.~A.} \emph{et~al.}
\newblock \bibinfo{title}{High, size-dependent quality factor in an array of
  graphene mechanical resonators}.
\newblock \emph{\bibinfo{journal}{Nano Lett.}} \textbf{\bibinfo{volume}{11}},
  \bibinfo{pages}{1232--1236} (\bibinfo{year}{2011}).

\bibitem{CartamilBueno2016}
\bibinfo{author}{Cartamil-Bueno, S.~J.} \emph{et~al.}
\newblock \bibinfo{title}{Colorimetry technique for scalable characterization
  of suspended graphene}.
\newblock \emph{\bibinfo{journal}{Nano Lett.}} \textbf{\bibinfo{volume}{16}},
  \bibinfo{pages}{6792--6796} (\bibinfo{year}{2016}).

\bibitem{Berger2016sheets}
\bibinfo{author}{Berger, C.~N.}, \bibinfo{author}{Dirschka, M.} \&
  \bibinfo{author}{Vijayaraghavan, A.}
\newblock \bibinfo{title}{Ultra-thin graphene{\textendash}polymer
  heterostructure membranes}.
\newblock \emph{\bibinfo{journal}{Nanoscale}} \textbf{\bibinfo{volume}{8}},
  \bibinfo{pages}{17928--17939} (\bibinfo{year}{2016}).

\bibitem{burningLin2011}
\bibinfo{author}{Lin, Y.-C.} \emph{et~al.}
\newblock \bibinfo{title}{Graphene annealing: How clean can it be?}
\newblock \emph{\bibinfo{journal}{Nano Lett.}} \textbf{\bibinfo{volume}{12}},
  \bibinfo{pages}{414--419} (\bibinfo{year}{2011}).

\bibitem{burningAhn2016}
\bibinfo{author}{Ahn, Y.}, \bibinfo{author}{Kim, J.},
  \bibinfo{author}{Ganorkar, S.}, \bibinfo{author}{Kim, Y.-H.} \&
  \bibinfo{author}{Kim, S.-I.}
\newblock \bibinfo{title}{Thermal annealing of graphene to remove polymer
  residues}.
\newblock \emph{\bibinfo{journal}{Mater. Express}}
  \textbf{\bibinfo{volume}{6}}, \bibinfo{pages}{69--76} (\bibinfo{year}{2016}).

\bibitem{burningHuang2014}
\bibinfo{author}{Huang, L.-W.} \emph{et~al.}
\newblock \bibinfo{title}{Characterization of the cleaning process on a
  transferred graphene}.
\newblock \emph{\bibinfo{journal}{J. Vac. Sci. Technol.}}
  \textbf{\bibinfo{volume}{32}}, \bibinfo{pages}{050601}
  (\bibinfo{year}{2014}).

\bibitem{burningGammelgaard2014}
\bibinfo{author}{Gammelgaard, L.} \emph{et~al.}
\newblock \bibinfo{title}{Graphene transport properties upon exposure to {PMMA}
  processing and heat treatments}.
\newblock \emph{\bibinfo{journal}{2D Mater.}} \textbf{\bibinfo{volume}{1}},
  \bibinfo{pages}{035005} (\bibinfo{year}{2014}).

\bibitem{residuesJain2018}
\bibinfo{author}{Jain, A.} \emph{et~al.}
\newblock \bibinfo{title}{Minimizing residues and strain in {2D} materials
  transferred from {PDMS}}.
\newblock \emph{\bibinfo{journal}{Nanotechnology}}
  \textbf{\bibinfo{volume}{29}}, \bibinfo{pages}{265203}
  (\bibinfo{year}{2018}).

\bibitem{RuizVargas2011}
\bibinfo{author}{Ruiz-Vargas, C.~S.} \emph{et~al.}
\newblock \bibinfo{title}{Softened elastic response and unzipping in chemical
  vapor deposition graphene membranes}.
\newblock \emph{\bibinfo{journal}{Nano Lett.}} \textbf{\bibinfo{volume}{11}},
  \bibinfo{pages}{2259--2263} (\bibinfo{year}{2011}).

\bibitem{CastellanosGomez2012}
\bibinfo{author}{Castellanos-Gomez, A.} \emph{et~al.}
\newblock \bibinfo{title}{Elastic properties of freely suspended {MoS}$_2$
  nanosheets}.
\newblock \emph{\bibinfo{journal}{Adv. Mater.}} \textbf{\bibinfo{volume}{24}},
  \bibinfo{pages}{772--775} (\bibinfo{year}{2012}).

\bibitem{Nicholl2015}
\bibinfo{author}{Nicholl, R.~J.} \emph{et~al.}
\newblock \bibinfo{title}{The effect of intrinsic crumpling on the mechanics of
  free-standing graphene}.
\newblock \emph{\bibinfo{journal}{Nat. Commun.}} \textbf{\bibinfo{volume}{6}}
  (\bibinfo{year}{2015}).

\bibitem{Li2015}
\bibinfo{author}{Li, Z.} \emph{et~al.}
\newblock \bibinfo{title}{Deformation of wrinkled graphene}.
\newblock \emph{\bibinfo{journal}{{ACS} Nano}} \textbf{\bibinfo{volume}{9}},
  \bibinfo{pages}{3917--3925} (\bibinfo{year}{2015}).

\bibitem{Suk2010}
\bibinfo{author}{Suk, J.~W.}, \bibinfo{author}{Piner, R.~D.},
  \bibinfo{author}{An, J.} \& \bibinfo{author}{Ruoff, R.~S.}
\newblock \bibinfo{title}{Mechanical properties of monolayer graphene oxide}.
\newblock \emph{\bibinfo{journal}{{ACS} Nano}} \textbf{\bibinfo{volume}{4}},
  \bibinfo{pages}{6557--6564} (\bibinfo{year}{2010}).

\bibitem{GmezNavarro2008}
\bibinfo{author}{G{\'{o}}mez-Navarro, C.}, \bibinfo{author}{Burghard, M.} \&
  \bibinfo{author}{Kern, K.}
\newblock \bibinfo{title}{Elastic properties of chemically derived single
  graphene sheets}.
\newblock \emph{\bibinfo{journal}{Nano Lett.}} \textbf{\bibinfo{volume}{8}},
  \bibinfo{pages}{2045--2049} (\bibinfo{year}{2008}).

\bibitem{Ferrari2007}
\bibinfo{author}{Ferrari, A.~C.}
\newblock \bibinfo{title}{Raman spectroscopy of graphene and graphite:
  Disorder, electron{\textendash}phonon coupling, doping and nonadiabatic
  effects}.
\newblock \emph{\bibinfo{journal}{Solid State Commun.}}
  \textbf{\bibinfo{volume}{143}}, \bibinfo{pages}{47--57}
  (\bibinfo{year}{2007}).

\bibitem{Kim2012}
\bibinfo{author}{Kim, K.} \emph{et~al.}
\newblock \bibinfo{title}{Raman spectroscopy study of rotated double-layer
  graphene: Misorientation-angle dependence of electronic structure}.
\newblock \emph{\bibinfo{journal}{Phys. Rev. Lett.}}
  \textbf{\bibinfo{volume}{108}} (\bibinfo{year}{2012}).

\bibitem{Ni2008}
\bibinfo{author}{Ni, Z.~H.} \emph{et~al.}
\newblock \bibinfo{title}{Tunable stress and controlled thickness modification
  in graphene by annealing}.
\newblock \emph{\bibinfo{journal}{{ACS} Nano}} \textbf{\bibinfo{volume}{2}},
  \bibinfo{pages}{1033--1039} (\bibinfo{year}{2008}).

\bibitem{Eckmann2012}
\bibinfo{author}{Eckmann, A.} \emph{et~al.}
\newblock \bibinfo{title}{Probing the nature of defects in graphene by raman
  spectroscopy}.
\newblock \emph{\bibinfo{journal}{Nano Lett.}} \textbf{\bibinfo{volume}{12}},
  \bibinfo{pages}{3925--3930} (\bibinfo{year}{2012}).

\bibitem{siskins2019}
\bibinfo{author}{{\v{S}}i{\v{s}}kins, M.} \emph{et~al.}
\newblock \bibinfo{title}{High-temperature electronic devices enabled by
  {hBN}-encapsulated graphene}.
\newblock \emph{\bibinfo{journal}{Appl. Phys. Lett.}}
  \textbf{\bibinfo{volume}{114}}, \bibinfo{pages}{123104}
  (\bibinfo{year}{2019}).

\bibitem{Son2017}
\bibinfo{author}{Son, S.-K.} \emph{et~al.}
\newblock \bibinfo{title}{Graphene hot-electron light bulb: incandescence from
  {hBN}-encapsulated graphene in air}.
\newblock \emph{\bibinfo{journal}{2D Mater.}} \textbf{\bibinfo{volume}{5}},
  \bibinfo{pages}{011006} (\bibinfo{year}{2017}).

\bibitem{bunch2008mechanical}
\bibinfo{author}{Bunch, J.~S.}
\newblock \emph{\bibinfo{title}{Mechanical and electrical properties of
  graphene sheets}} (\bibinfo{publisher}{Cornell University Ithaca, NY},
  \bibinfo{year}{2008}).

\bibitem{qcPonomarenko2010}
\bibinfo{author}{Ponomarenko, L.~A.} \emph{et~al.}
\newblock \bibinfo{title}{Density of states and zero landau level probed
  through capacitance of graphene}.
\newblock \emph{\bibinfo{journal}{Phys. Rev. Lett.}}
  \textbf{\bibinfo{volume}{105}} (\bibinfo{year}{2010}).

\bibitem{qcXia2009}
\bibinfo{author}{Xia, J.}, \bibinfo{author}{Chen, F.}, \bibinfo{author}{Li, J.}
  \& \bibinfo{author}{Tao, N.}
\newblock \bibinfo{title}{Measurement of the quantum capacitance of graphene}.
\newblock \emph{\bibinfo{journal}{Nature Nanotech.}}
  \textbf{\bibinfo{volume}{4}}, \bibinfo{pages}{505--509}
  (\bibinfo{year}{2009}).

\end{thebibliography}

\end{document}